\documentclass[%
 reprint,
 amsmath,amssymb,
 aps,onecolumn,nofootinbib,
]{revtex4-2}
\usepackage{multirow}
\usepackage{graphicx}
\usepackage{dcolumn}
\usepackage{bm}
\usepackage{fancyhdr}
\usepackage[colorlinks]{hyperref}
\usepackage[nameinlink,noabbrev]{cleveref}
\hypersetup{
    colorlinks=true,
    linkcolor=blue,
    filecolor=magenta,      
    urlcolor=blue,
   citecolor=blue
}
\usepackage{graphicx}
\usepackage{dcolumn}
\usepackage{bm}
\usepackage{scalerel}
\usepackage{tikz}
\usetikzlibrary{svg.path}

\definecolor{orcidlogocol}{HTML}{A6CE39}
\tikzset{
  orcidlogo/.pic={
    \fill[orcidlogocol] svg{M256,128c0,70.7-57.3,128-128,128C57.3,256,0,198.7,0,128C0,57.3,57.3,0,128,0C198.7,0,256,57.3,256,128z};
    \fill[white] svg{M86.3,186.2H70.9V79.1h15.4v48.4V186.2z}
                 svg{M108.9,79.1h41.6c39.6,0,57,28.3,57,53.6c0,27.5-21.5,53.6-56.8,53.6h-41.8V79.1z M124.3,172.4h24.5c34.9,0,42.9-26.5,42.9-39.7c0-21.5-13.7-39.7-43.7-39.7h-23.7V172.4z}
                 svg{M88.7,56.8c0,5.5-4.5,10.1-10.1,10.1c-5.6,0-10.1-4.6-10.1-10.1c0-5.6,4.5-10.1,10.1-10.1C84.2,46.7,88.7,51.3,88.7,56.8z};
  }
}

\newcommand\orcidicon[1]{\href{https://orcid.org/#1}{\mbox{\scalerel*{
\begin{tikzpicture}[yscale=-1,transform shape]
\pic{orcidlogo};
\end{tikzpicture}
}{|}}}}
\usepackage{amsmath}
\usepackage{amssymb, amsfonts}
\usepackage[greek,english]{babel}
\begin{document}

\preprint{APS/123-QED}

\title{Reaction Rates in Quasiequilibrium States}
%


\author{Kamel Ourabah \orcidicon{0000-0003-0515-6728},}\email{kam.ourabah@gmail.com, kourabah@usthb.dz}
\address{Theoretical Physics Laboratory, Faculty of Physics, University of Bab-Ezzouar, USTHB, Boite Postale 32, El Alia, Algiers 16111, Algeria}

\date{\today}

\begin{abstract}
Non-Maxwellian distributions are commonly observed across a wide range of systems and scales. While direct observations provide the strongest evidence for these distributions, they also manifest indirectly through their influence on processes and quantities that strongly depend on the energy distribution, such as reaction rates. In this paper, we investigate reaction rates in the general context of quasiequilibrium systems, which exhibit only local equilibrium. The hierarchical structure of these systems allows their statistical properties to be represented as a superposition of statistics, i.e., superstatistics. Focusing on the three universality classes of superstatistics—$\chi^2$, inverse-$\chi^2$, and log-normal—we examine how these nonequilibrium distributions influence reaction rates. We analyze, both analytically and numerically, reaction rates for processes involving tunneling phenomena, such as fusion, and identify conditions under which quasiequilibrium distributions outperform Maxwellian distributions in enhancing fusion reactivities. To provide a more detailed quantitative analysis, we further employ semi-empirical cross sections to evaluate the effect of these nonequilibrium distributions on ionization and recombination rates in a plasma.
\end{abstract}

\maketitle


\section{Introduction}
\label{1}

A naive application of statistical mechanics might suggest that the universe should be dominated by matter in equilibrium, characterized by the Maxwell–Boltzmann distribution.
Yet, the environments we observe throughout the universe tell a different story. Non-Maxwellian distributions are commonly observed across a wide range of systems, spanning nearly all scales—from galaxy clusters to the solar neighborhood, and all the way down to Earth-based experiments. More specifically, non-Maxwellian distributions are ubiquitous in space environments, frequently manifesting in space plasmas \cite{azerty1,azerty2,azerty3,azerty4,azerty5} as well as in gravity-dominated systems such as stellar and galaxy clusters \cite{Astro1,Astro2,Astro3,Astro4,galaxy}. In laboratory settings, these distributions are routinely observed across a broad range of experimental setups, including plasmas \cite{plasma1,Labplasma1,Labplasma2}, cold atoms \cite{Cold1,Cold2}, trapped ions \cite{Devoe}, granular gases \cite{gran}, spin glasses \cite{spin}, particles in active baths \cite{bath}, graphene membranes \cite{graphene}, cell monolayers \cite{cell}, and high-energy collisional experiments \cite{HE1,HE2}.

This observation, on its own, is not particularly surprising and can be explained on physical grounds. In fact, the key process driving systems toward Maxwell–Boltzmann equilibrium is two-body interparticle collisions. Still, in a variety of scenarios, this mechanism fails to bring the system to equilibrium within a reasonable timescale.
In plasmas, for example, collisions typically occur over a timescale much longer than that associated with the evolution of the mean electric and magnetic fields generated by the plasma. In self-gravitating systems, the situation is even worse, as the relaxation time diverges approximately linearly with the number of particles, meaning that for sufficiently massive objects, it can exceed the age of the universe \cite{Chandra1,Chandra2}. In these conditions, the system remains trapped in a quasiequilibrium state.

Determining the steady state of a nonequilibrium system is a very complex task, as it fundamentally requires knowledge of the complete history of perturbations the system has experienced. Nonetheless, despite this complexity, the distributions we observe frequently display universal features and can be categorized into a limited number of universality classes. Arguably, the most universal approach for representing these distributions is the concept of superstatistics \cite{super0} (see also Refs. \cite{sa1,sa2,sa3,sa4,Z4} for earlier works in the same spirit, and Ref. \cite{hyper} for the related notion of hyperensembles). In a nutshell, this approach explains the emergence of these distributions from spatio-temporal fluctuations of the temperature or other intensive parameters. As a result, it allows for the reproduction of the characteristic traits of the most commonly observed distributions in various situations of physical interest. Recent research provides robust and well-documented evidence for superstatistics across various physical contexts, particularly in turbulence \cite{Tur1,stur2,Tur2}, plasmas \cite{Z4,Plasma1,Plasma2,splasma3}, cold atoms \cite{Rouse}, self-gravitating systems \cite{OurabahPRD,OurabahPRE}, and high-energy physics \cite{sHE1,sHE2}. Furthermore, the concept has proven effective in modeling a range of scenarios beyond physics, including traffic dynamics \cite{traffic}, finance \cite{Market0,Market}, power grid fluctuations \cite{nature}, DNA architecture \cite{ss2,ss22}, rainfall statistics \cite{ss3}, air pollution \cite{ss4}, and cognitive processes \cite{ss5}, among others.

The strongest evidence for non-Maxwellian distributions naturally comes from direct observation, whenever possible. Nonetheless, they also manifest indirectly, as they are hidden behind very complex processes that strongly depend on the specific form of the distribution. A notable example is the reaction rate in two-body collisions, which is very sensitive to specific details and highly influenced by the energy distribution of the reactants. Accurately determining reaction rates—whether for fusion reactions or ionization and recombination in plasmas—is essential. In astrophysics for example, these rates significantly impact the changes in the abundances of various species within astrophysical environments. Precise rate determination is also crucial for controlled nuclear fusion, which has attracted significant interest in light of global warming concerns, with projects like ITER currently in progress and DEMO anticipated for mid-century.

Reaction rates have traditionally been studied using Maxwellian distributions, which are well-suited for equilibrium or near-equilibrium conditions. Recently, however, the influence of non-Maxwellian distributions has gained attention at both theoretical and experimental levels. From an experimental standpoint, recent high-sensitivity measurements of reaction rates have been conducted in ion traps, where radio-frequency heating generates a high-energy tail in the ion distribution \cite{Wild2023}. From a theoretical perspective, previous studies have examined the effects of specific forms of non-Maxwellian distributions such as kappa distributions and bimodal Maxwellian distributions \cite{Onofrio0,Onofrio1}, as well as (quasi)-Maxwellian distributions with anisotropic temperatures \cite{Tok1,Tok2,Tok3,Tok4,Tok5}. Simulations in the latter case indicate that fusion reactivity could potentially double under ITER-like conditions \cite{Tok3}.

In this paper, we aim to take a broader perspective on reaction rates within the general context of a quasiequilibrium system experiencing temperature fluctuations. Fundamentally, the only assumption is the existence of different spatio-temporal scales, which allow enough time for the system to relax to a local equilibrium state and to remain there for a phenomenologically significant period. While our focus will be particularly on the three universality classes of superstatistics commonly observed in many experimentally relevant situations, the insights we present can be readily extended to encompass other scenarios, provided there is sufficient separation between the spatio-temporal scales of the relevant dynamics for superstatistics to apply.

The rest of the paper is organized as follows. In Sec. \ref{SecII}, we provide a detailed overview of the concept of superstatistics, focusing on the three universality classes that have shown considerable empirical validation across different studies. We validate these classes by comparing them with simulation data from ion–molecule reaction experiments. In Sec. \ref{SecIII}, we analyze their impact on reaction rates in an idealized prototype model that simulates tunneling phenomena, such as fusion. In Sec. \ref{SecIV}, we investigate reaction rates in a more realistic context, using semi-empirical cross-section formulas to examine their influence on ionization and recombination rates in plasmas. We conclude in Sec. \ref{SecV} and suggest a few avenues for future exploration.

\section{Superstatistics and nonequilibrium stationary distributions}\label{SecII}

To provide context, let us clarify the overall situation we are addressing. We consider the general scenario of a system that is not in full thermodynamic equilibrium but only exhibits \textit{local} equilibrium. In this context, the system can be viewed as consisting of small subsystems, or cells, each in local equilibrium. These subsystems act as regions from which thermalization gradually spreads throughout the entire system. Since the relaxation time $\tau_0$ of a cell is much shorter than the relaxation time $\tau$ of the entire system (at least when long-range interactions are not involved), there exists an intermediate timescale $t$ such that
\begin{equation}
\tau_0 \ll t \ll \tau,  
\end{equation}
on which the cells are nearly in equilibrium, while the overall system has not yet reached full equilibrium. Under these conditions, the system is said to be in a \textit{quasiequilibrium} state \cite{Mishin2015,Hickman2016} or a \textit{quasistationary} state (see e.g. Landau and Lifschitz \cite{Landau}). Such systems, exhibiting hierarchical structures in their underlying dynamics, can be described across two (or more) distinct scales. At the small scale, the system has reached equilibrium, with its local statistical properties governed by Boltzmann statistics. This is characterized by a well-defined local inverse temperature, $\beta \equiv 1/T$ (we set the Boltzmann constant $k_B$ to unity throughout this paper). The local probability of the system being in a state with energy $\varepsilon$ is given by $p(E | \beta) \propto \exp(-\beta E)$. However, at a larger scale, the inverse temperature $\beta$ fluctuates among different cells. When these fluctuations take place over a time scale much longer than the typical relaxation time of the local dynamics $\tau_0$, a distribution $f(\beta)$ can be used to characterize the variation of $\beta$\footnote{In general, one can consider fluctuations of any other intensive parameter that varies over a timescale much longer than the relaxation time of the local dynamics, such as the chemical potential \cite{sHE1}, the energy dissipation rate \cite{beta1}, or volatility in econophysics \cite{beta2}. From a mathematical perspective, this scale separation makes superstatistics a form of slow modulation \cite{Allegrini}.}. This leads to the overall statistical properties of the system being governed by

\begin{equation}\label{B}
p(E) = \int_{0}^{\infty} d\beta \, f(\beta) \frac{\exp(-\beta E)}{Z(\beta)},
\end{equation}
where $Z(\beta)$ is the (local) partition function\footnote{An alternative to Eq. (\ref{B}) consists of omitting $Z(\beta)$ and normalizing the resulting distribution afterward. These are known as Type-A and Type-B superstatistics (see e.g. \cite{beta1}).}. Note that $p(E)$ is simply the Laplace transform of the (rescaled) temperature distribution $\tilde{f}(\beta) := f(\beta) / Z(\beta)$.

While the superstatistics approach theoretically accommodates any probability density function (PDF) $f(\beta)$, in practice, only a limited number of relevant distributions are typically observed. Specifically, three universality classes often emerge, namely (a) 
$\chi^2$ superstatistics (which leads to Tsallis thermostatistics \cite{super0}); (b) inverse-$\chi^2$ superstatistics; and (c) log-normal superstatistics. The existence of these classes can be attributed to three typical phenomenological scenarios that are likely responsible for the emergence of the random variable \cite{CLT}, and they can also be understood through the maximum entropy principle \cite{MAXENT}. Given the strong empirical support for these classes and their transparent statistical origin, our analysis will center on them. The three classes are as follows:


\begin{itemize}
    \item[(a)] \textbf{$\chi^2$ superstatistics:} In this case, the inverse temperature $\beta$ follows a $\chi^2$ distribution with degree $n$,
    \begin{equation}\label{f1}
    f(\beta) = \frac{1}{\Gamma\left(\frac{n}{2}\right)} \left(\frac{n}{2\beta_0}\right)^{n/2} \beta^{n/2-1} \exp\left(-\frac{n\beta}{2\beta_0}\right),
    \end{equation}
    where $\beta_0 \equiv \langle \beta \rangle$ is the average value of $\beta$. Assuming a ($d$-dimensional) Maxwellian distribution within the cells, we can derive the emergent velocity distribution from Eq. (\ref{B}) as follows:
    \begin{equation}\label{B1}
    P(v) = \int_{0}^{\infty} d\beta f(\beta) \left(\frac{m}{2\pi \beta}\right)^{d/2} \exp\left(-\frac{\beta mv^2}{2}\right)
    = \frac{\beta_0 m}{\pi^{d/2} \Gamma\left(\frac{n}{2}\right)} \frac{\Gamma\left(\frac{n+d}{2}\right)}{(1 + \frac{\beta_0 m v^2}{n})^{(n+d)/2}}.
    \end{equation}
    This corresponds to the distributions encountered in Tsallis thermostatistics ($q$-Gaussian). This can be made more apparent by introducing an entropic index $\tilde{q}$ and an effective inverse temperature $\tilde{\beta}$, defined as follows:

\begin{equation}
\tilde{q} \equiv 1 + \frac{2}{n+d} \quad \text{and} \quad  \tilde{\beta} \equiv \left ( \frac{n+d}{n} \right )
 \beta_0.   
\end{equation}
In this case, Eq. (\ref{B1}) can be reexpressed in the language of Tsallis thermostatistics as     

\begin{equation}
P(v) \propto \left( 1 + (\tilde{q} - 1) \frac{\tilde{\beta} mv^2}{2} \right)^{\frac{1}{1 - \tilde{q}}}.
\end{equation}
   
It is also identical to the so-called kappa distribution used in plasma physics, in the form advocated by Leubner \cite{Leubner} and adopted by others \cite{Liva1,plasma2,Shizgal}, although it slightly differs from the original kappa distribution introduced by Vasyliunas in the late 1960s \cite{azerty2}.
    
Note that, for large values of $|v|$, Eq. (\ref{B1}) exhibits a power-law behavior. This characteristic makes it highly effective for modeling a wide range of systems where suprathermal tails often follow a power-law decay. Such quasi-power-law distributions are frequently observed in various physical settings, including plasmas \cite{plasma1,plasma2}, cold atoms \cite{cold1,cold2}, high-energy collisions \cite{HE1,HE2}, and self-gravitating systems \cite{Astro1,galaxy}.

    \item[(b)] \textbf{Inverse-$\chi^2$ superstatistics:} Here, the temperature itself $(\beta^{-1})$, instead of $\beta$, is assumed to follow a $\chi^2$ distribution, leading to an inverse-$\chi^2$ distribution for $\beta$,
    \begin{equation}\label{f2}
    f(\beta) = \frac{\beta_0}{\Gamma\left(\frac{n}{2}\right)} \left(\frac{n\beta_0}{2}\right)^{n/2} \beta^{-(n/2 + 2)} \exp\left(-\frac{n\beta_0}{2\beta}\right).
    \end{equation}
    The corresponding velocity distribution in this case can be expressed as
    \begin{equation}\label{B2}
    P(v) = \frac{2\beta_0}{\Gamma\left(\frac{n}{2}\right)} \left(\frac{m}{2\pi}\right)^{d/2} \left(\frac{n}{2\beta_0}\right)^{n/2} \left(\frac{mv^2}{\beta_0 n}\right)^{(n-d)/4} \mathcal{K}_{\frac{2-d+n}{2}}\left(\sqrt{nm\beta_0}|v|\right),
    \end{equation}
    where $\mathcal{K}_{\alpha}(x)$ is the modified Bessel function of the second kind. For large velocities, Eq. (\ref{B2}) produces exponential tails. This kind of exponential decay has been identified in various nonequilibrium systems, including vortex glasses/liquids \cite{VG}, fusion plasmas \cite{Z4}, and harmonic oscillators interacting with solvent baths \cite{invchi11}. Similar patterns have also been observed beyond physics, for example, in cancer-specific mortality rate distributions \cite{invchi22}.
    
    \item[(c)] \textbf{Log-normal superstatistics:} In this scenario, $\beta$ follows a log-normal distribution,
    \begin{equation}\label{f3}
    f(\beta) = \frac{1}{\sqrt{2\pi} s \beta} \exp\left(-\frac{(\ln \beta - \mu)^2}{2s^2}\right),
    \end{equation}
    with the average value of $\beta$ given by $\beta_0 = \mu e^{s^2/2}$. In this case, the corresponding velocity distribution $B(v)$ does not have a closed-form expression but can be easily computed numerically. Log-normal superstatistics commonly appears in complex systems with cascading dynamics, such as turbulence \cite{stur2,stur3}. It also effectively captures observed distributions in space plasmas \cite{splasma3} and gravitational systems \cite{OurabahPRE}, and has been identified in other contexts beyond physics \cite{nature,Xu,Jizba2}.
    
\end{itemize}

\begin{figure*}
\centering
\begin{minipage}[t]{0.3\linewidth}
\includegraphics[width=1\linewidth]{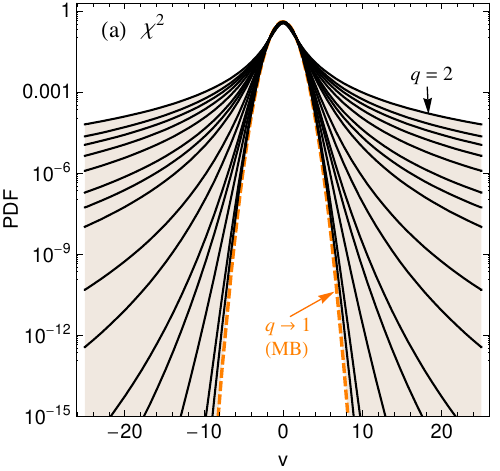}
\end{minipage}
\begin{minipage}[t]{0.3\linewidth}
\includegraphics[width=1\linewidth]{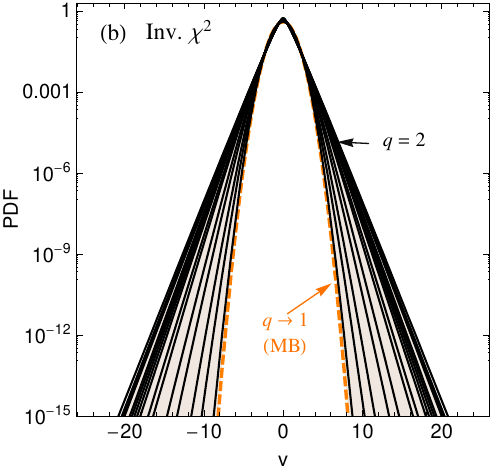}
\end{minipage}
\begin{minipage}[t]{0.3\linewidth}
\includegraphics[width=1\linewidth]{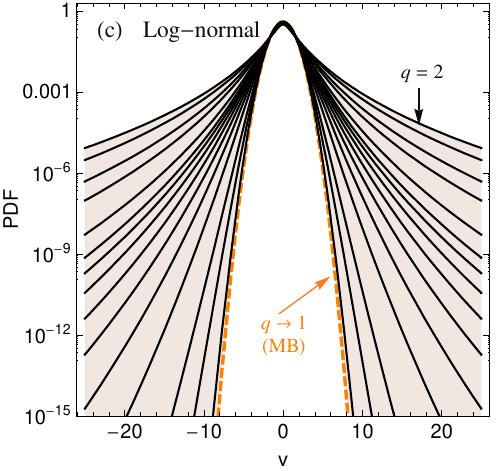}
\end{minipage}
\caption{Examples of one-dimensional superstatistical velocity PDFs $P(v)$ for (a) $\chi^2$, (b) inverse-$\chi^2$, and (c) log-normal superstatistics, plotted for different values of $q \equiv \langle \beta^2 \rangle / \beta_0^2$ [cf. Eq. (\ref{q})], on a logarithmic scale to emphasize the distribution tails. The parameter $\beta_0 m$ is set to 1.} \label{Fig1}
\end{figure*}

One may note that the asymptotic characteristics of these distributions reflect the most typical behaviors observed: For large $|v|$, $\chi^2$ superstatistics produces power-law tails, inverse-$\chi^2$ superstatistics exhibits exponential decay, and log-normal superstatistics leads to truncated power laws. This diversity offers a rich foundation for modeling non-Maxwellian distributions commonly seen in nonequilibrium settings. On the other hand, when the amplitude of the fluctuations is small, it can be shown that all classes converge to a universal behavior \cite{super0}
\begin{equation}\label{al}
P(v) \propto\left(1+{\frac{(q-1)}{8} \beta_0^2 m^2 v^4}\right) e^{-\frac{\beta_0 m v^2}{2}},
\end{equation}
where $q:= \langle \beta^2 \rangle / \beta_0^2$. Such distributions, which involve a quadratic correction beyond the equilibrium distribution, are also often encountered in plasma physics.
Originally introduced to explain solitary electrostatic structures with density depletions in the upper ionosphere, these distributions are known in the plasma physics community as nonthermal or Cairns distributions \cite{Cairns1,Cairns2}.

In what follows, we will restrict ourselves to the three universality classes of superstatistics and their universal limit (\ref{alpha}). However, other types of distributions may emerge under specific conditions in simpler scenarios. For example, bimodal Maxwellian distributions, recently studied in the context of fusion reactivity \cite{Onofrio1}, are given by

\begin{equation}
P(v)=x_1 P_{M B}\left(v, \beta_1\right)+x_2 P_{M B}\left(v, \beta_2\right),
\end{equation}
where $P_{MB}$ denotes the Maxwellian distribution and $\beta_1$ and $\beta_2$ represent different inverse temperatures. These bi-Maxwellians can also be mapped into superstatistics by assuming a two-level distribution for $f(\beta)$, namely
\begin{equation}
f(\beta)= x_1 \delta (\beta - \beta_1) + x_2 \delta (\beta - \beta_2),
\end{equation}
with $x_1+x_2=1$. Another potentially interesting scenario is that of stretched exponential distributions, which naturally arise in granular gases \cite{gran} and in plasma physics under specific conditions\footnote{Stretched exponentials, i.e., $P(v) \propto \exp\left(- \alpha |v|^m\right)$, are encountered in various plasma-physics scenarios. For instance, in turbulent plasmas, values of $m$ typically range from 3.6 to 5 \cite{sp1,sp2}. Furthermore, it has been demonstrated that in laser-produced plasmas, inverse bremsstrahlung heating yields a stretched exponential distribution with $m = 5$ \cite{sp3}. The implications of these distributions on reaction rates are discussed in Ref. \cite{PRA}.}, and can also be interpreted as a form of superstatistics (see Ref. \cite{Beck2006} for details).

Figure \ref{Fig1} presents examples of (one-dimensional) velocity distributions arising from the three universality classes: $\chi^2$ [Eq. (\ref{B1})], inverse-$\chi^2$ [Eq. (\ref{B2})], and log-normal (computed numerically). To facilitate comparisons among these classes, we parameterize the distributions using the parameter $q := {\langle \beta^2 \rangle}/{\beta_0^2}$. For the three universality classes, $q$ can be expressed in terms of the parameters of $f(\beta)$ as follows:

\begin{equation}\label{q}
\begin{aligned}
q \equiv & \frac{\langle \beta^2 \rangle_{\chi^2}}{\beta_0^2} = 1 + \frac{2}{n} \quad (n > 2), \\
q \equiv & \frac{\langle \beta^2 \rangle_{\text{inv-}\chi^2}}{\beta_0^2} = \frac{n}{n-2}, \\
q \equiv & \frac{\langle \beta^2 \rangle_{LN}}{\beta_0^2} = e^{s^2}.
\end{aligned}
\end{equation}
This parameterization is useful as it facilitates a fair comparison among the different classes in terms of fluctuation strength. In fact, a larger deviation of $q$ from unity indicates increased fluctuations. Conversely, when $q \to 1$, the variance of $f(\beta)$ tends to zero (i.e., $f(\beta)$ collapses into a Dirac delta distribution), and equilibrium Maxwellian distributions are restored, independent of the superstatistics class.

\begin{figure*}
\centering
\begin{minipage}[t]{0.4\linewidth}
\includegraphics[width=1\linewidth]{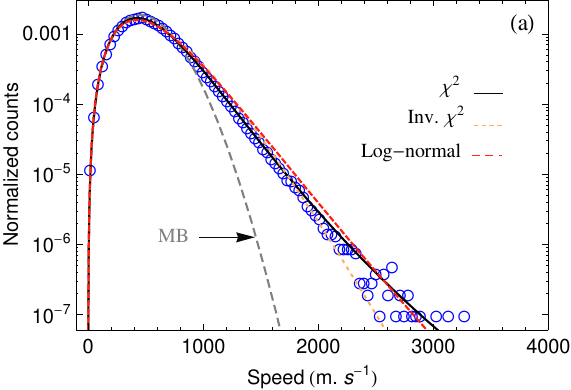}
\end{minipage}
\begin{minipage}[t]{0.4\linewidth}
\includegraphics[width=1\linewidth]{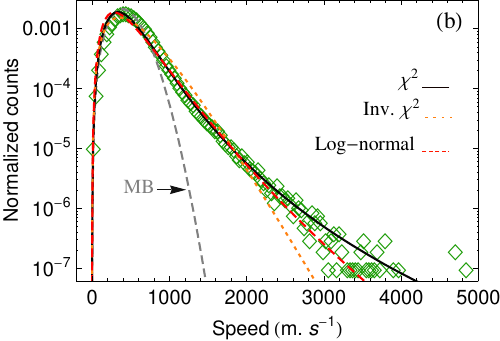}
\end{minipage}
\caption{Simulated (relative) velocity distributions of ions in a Maxwellian hydrogen gas background, derived from ion–molecule reaction experiments as discussed in Ref. \cite{Wild2023}. These distributions are fitted with the three universality classes of superstatistics, namely $\chi^2$ (solid line), inverse-$\chi^2$ (dotted line), and log-normal (dashed line). The Maxwellian distribution is shown for reference.} \label{Fig2}
\end{figure*}

To illustrate the relevance of these distributions in experimentally relevant situations, Figure \ref{Fig2} presents a comparison with simulated (relative) velocity distributions of ions and equilibrium hydrogen gas, based on the ion–molecule reaction experiments described in Ref. \cite{Wild2023}. One may see that the three universality classes effectively capture the overall trends and provide a good fit to the data. For reference, the Maxwellian distribution is also shown for comparison.


At this point, it is important to note that, when evaluating reaction rates, the primary focus should be on the energy distribution of the reactants. Since the average energy is conserved, comparisons between different distributions must be based on the same average energy, rather than temperature, which is not constant in this context. In our case, the average energy is calculated by combining the moments of the Maxwellian distribution with those of $f(\beta)$. For the three universality classes, the moments of $f(\beta)$ are given by\footnote{As superstatistical distributions are essentially Maxwellian distributions averaged over $f(\beta)$, the velocity moments can be written as $\left\langle v^l\right\rangle \equiv \int v^l P(v) d v=\left\langle\left\langle v^l\right\rangle_{\mathrm{MB}}\right\rangle_{f}$, where $\langle\cdot\rangle_{\mathrm{MB}}$ denotes the average over the Maxwellian velocity distribution and 
$\langle\cdot\rangle_{{f}}$  represents the average over $f(\beta)$.}, namely

\begin{equation}\label{moments}
\begin{aligned}
\left\langle\beta^l\right\rangle_{\chi^2} & =\frac{\Gamma\left(\frac{n}{2}+l\right)}{\Gamma\left(\frac{n}{2}\right)}\left(\frac{2}{n}\right)^l \beta_0^l, \\
\left\langle\beta^l\right\rangle_{\text {inv- } \chi^2} & =\frac{\Gamma\left(\frac{n}{2}+1-l\right)}{\Gamma\left(\frac{n}{2}\right)}\left(\frac{n}{2}\right)^{l-1} \beta_0^l, \\
\left\langle\beta^l\right\rangle_{\mathrm{LN}} & =e^{l(l-1) s^2 / 2} \beta_0^l.
\end{aligned}
\end{equation}
In one dimension ($d=1$), one has for the three universality classes 
\begin{equation}\label{EE}
\langle E \rangle \equiv \frac{m \langle v^2 \rangle}{2} =  \frac{\phi_i(q)}{2 \beta_0} \quad (i = 1, 2, 3),
\end{equation}
where the auxiliary functions $\phi_i(q) \geq 1$ are defined as follows:
\begin{equation}\label{phi}
\begin{aligned}
\phi_1(q) \equiv & \frac{1}{2-q} \quad (1 < q < 2),\\
\phi_2(q) \equiv & \frac{2q-1}{q}, \\
\phi_3(q) \equiv & q,
\end{aligned}
\end{equation}
with $i = 1, 2,$ and $3$ corresponding, respectively, to $\chi^2$, inverse-$\chi^2$, and log-normal superstatistics. In the limit $q \to 1$, one has $\phi_i = 1$, and the Maxwellian average energy $\langle E \rangle = {1}/{2 \beta_0}$ is recovered, where $\beta_0$ represents the constant inverse temperature specific to the Maxwellian case.

In order to compare distributions based on the same average energy, we use Eqs. (\ref{q})-(\ref{phi}), to rewrite each superstatisrical distribution in terms of $\langle E \rangle$ and $q$. In the one-dimensional case, applying this to the $\chi^2$ class [i.e., Eq. (\ref{B1})] results in

\begin{equation}\label{p1}
p_{1}(E ; \langle E \rangle, q)= \mathcal{A}_1   \sqrt{\frac{1}{\langle E \rangle E}} \left(\frac{2}{ q-1} + \frac{E/ \langle E \rangle}{2  -  q}\right)^{\frac{1 + q}{2 - 2 q}},
\end{equation}
where the normalization constant reads as
\begin{equation}
\mathcal{A}_1= \frac{2^{\frac{1}{ q-1}} \left(\frac{1}{ q-1}\right)^{\frac{1}{ q-1}} \sqrt{\frac{1}{  \pi(2 -  q)}}  \Gamma\left(\frac{1}{2} + \frac{1}{ q-1}\right)}{\Gamma\left(\frac{1}{ q-1}\right)}.
\end{equation}

In the high-energy limit, the distribution (\ref{p1}) asymptotically follows a power law, $E^{-\alpha}$, where $\alpha = q/(q-1)$. This behavior is similar to that of the kappa distribution, examined in \cite{Onofrio1} for fusion reactivity, which also decays as a power law with $\alpha = \kappa + 3/2$.

For inverse-$\chi^2$ superstatistics [i.e., Eq. (\ref{B2})], the corresponding one-dimensional energy distribution reads as

\begin{equation}
p_{2}(E ; \langle E \rangle, q)= \mathcal{A}_2 \sqrt{E} \left(\frac{ ( q-1) \langle E \rangle E}{ 2q-1}\right)^{-\frac{-3 + q}{4 ( q-1)}} \mathcal{K}_{\frac{ 3q-1}{2q-2}}\left( \frac{\sqrt{ (1 - 3q + 2q^2) E/\langle E \rangle}}{ ( q-1)}\right),  
\end{equation}
where the normalization constant is given by
\begin{equation}
\mathcal{A}_2= \frac{2^{-\frac{1}{2} - \frac{q}{ q-1}} ( q-1) \left(\frac{2q-1}{ (q-1) \langle E \rangle}\right)^{\frac{q}{ q-1}} }{\sqrt{\pi} q \Gamma\left(\frac{q}{q-1}\right)}.
\end{equation}

For log-normal superstatistics, where a closed-form expression for the distribution $p_{3}(E ; \langle E \rangle, q)$ does not exist, the constraint of a constant average energy can be easily implemented numerically. In the limit $q \to 1$, the three superstatistical distributions converge to the Maxwellian distribution with the same average energy $\langle E \rangle$, namely 

\begin{equation}
p_{\mathrm{MB}}(E ; \langle E \rangle)=\sqrt{\frac{1}{2 \pi E \langle E \rangle }}  \exp (- E/ 2 \langle E \rangle).
\end{equation}

\begin{figure*}
\centering
\begin{minipage}[t]{0.3\linewidth}
\includegraphics[width=1\linewidth]{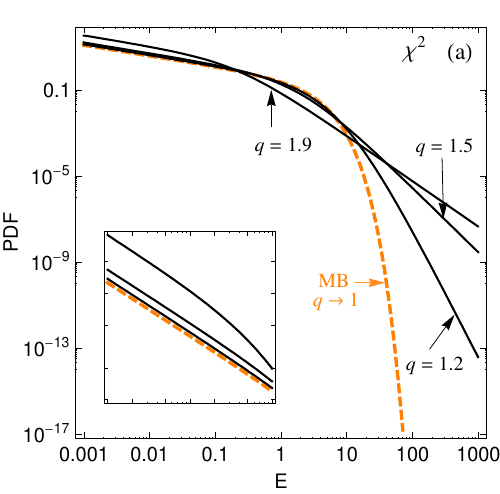}
\end{minipage}
\begin{minipage}[t]{0.3\linewidth}
\includegraphics[width=1\linewidth]{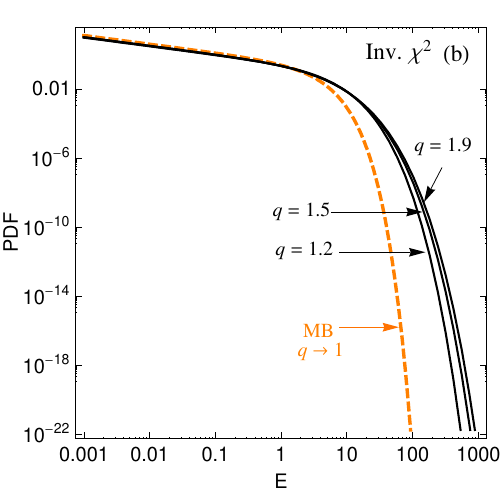}
\end{minipage}
\begin{minipage}[t]{0.3\linewidth}
\includegraphics[width=1\linewidth]{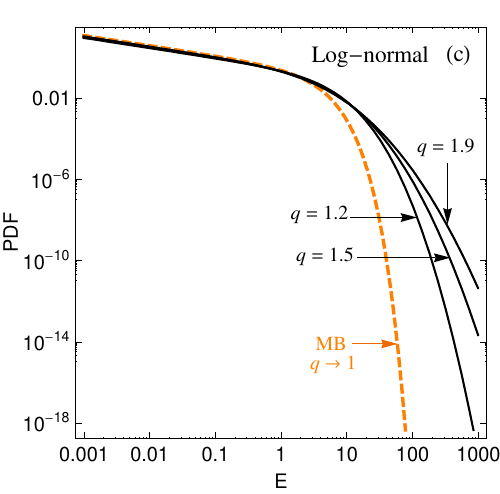}
\end{minipage}
\caption{Examples of superstatistical energy PDFs, $p(E)$, for (a) $\chi^2$, (b) inverse-$\chi^2$, and (c) log-normal superstatistics, all with the same average energy $\langle E \rangle$. The PDFs are plotted against energy $E$ (in arbitrary units) for different values of $q := \langle \beta^2 \rangle / \beta_0^2$.} \label{Fig3}
\end{figure*}

Figure \ref{Fig3} shows the energy distributions of the three universality classes of superstatistics, for a fixed average energy, across different values of  
$q:= \langle \beta^2\rangle/ \beta_0^2$. It is worth mentioning that these distributions not only have notable high-energy tails, but also more pronounced peaks at zero energy (or velocity) compared to the Maxwellian distribution. This is particularly evident for $\chi^2$ superstatistics (see inset), although it holds for all three universality classes. In fact, since these distributions are essentially Maxwellian distributions averaged over $f(\beta)$, they peak at $E = 0$. To quantify the degree of overpopulation at the origin, we define the following ratio:

\begin{equation}
\eta_i(q) := \frac{p_i(E=0 ; \langle E \rangle, q)}{p_{\mathrm{MB}}(E=0 ; \langle E \rangle)} \quad (i = 1, 2, 3),
\end{equation}
which can be easily computed using the moments of $f(\beta)$ [i.e., Eq. (\ref{moments})]. It reads as

\begin{equation}\label{alpha}
\begin{aligned}
\eta_1(q) \equiv & \frac{\sqrt{\frac{1}{ q-1}} \Gamma\left( \frac{1}{ q-1} -\frac{1}{2}\right)}{(2 - q) \Gamma\left(\frac{1}{ q-1}\right)}
 \quad (1 < q < 2),\\
\eta_2(q) \equiv & \frac{\left( q-1\right)^{\frac{3}{2}} \left( 2 q-1\right) \Gamma\left(\frac{3}{2} + \frac{q}{ q-1}\right)}{q \Gamma\left(\frac{q}{ q-1}\right)}
, \\
\eta_3(q) \equiv & q^{11/8},
\end{aligned}
\end{equation}
which is always above unity, indicating an overpopulation at the origin, and approaches unity as $q$ approaches 1.

\section{Reaction rates for tunneling phenomena: a prototype model}\label{SecIII}

The reaction rate per particle pair, or \textit{reactivity}, is determined by integrating the interaction cross sections, $\sigma$, with the relevant energy distribution of the interacting particles in the system. That is,
\begin{equation}\label{r}
 R \equiv   \langle\sigma v\rangle  =\int_0^{+\infty} P(v) \sigma v d v.
\end{equation}
Note that, although the integration extends from zero to infinity, the primary contributions to the integral (\ref{r}) arise from a narrow energy range\footnote{For
charged-particle reactions, this energy range is known as the Gamow window. This range is important for both nuclear experimentalists and theoreticians, as it identifies the window within which reaction cross sections have to be known.}, which depends on the specific energy dependence of the cross sections $\sigma(E)$ and the distribution $P(v)$. This is what makes reactivities very sensitive to the details. In this section, we aim to explore how the quasiequilibrium energy distributions discussed in Sec. \ref{SecII} affect the resulting average reactivity and identify the conditions under which they offer an enhancement over the Maxwellian case. More precisely, we examine reaction rates related to tunneling processes in an idealized one-dimensional scenario. In this context, the cross section can be expressed as

\begin{equation}
\sigma=\frac{\pi}{k^2} T(k)=\frac{\pi \hbar^2}{2 m E} T(E),
\end{equation}
where $k = \sqrt{2mE} / \hbar$ stands for the wave vector, and the particle velocity is expressed as $v = \hbar k / m$. The average reactivity is then determined as

\begin{equation}\label{R}
R_i \equiv 
\langle\sigma v\rangle_i  
 =\frac{\pi \hbar^2}{\sqrt{2} m^{3 / 2}} \int_0^{+\infty} p_i(E) T(E) E^{-1 / 2} d E,
\end{equation}
where $p_i(E)$ ($i=1,2,3$) represents one of the energy distributions discussed in Sec. \ref{SecII}, corresponding to one of the three universality classes of superstatistics. To assess the influence of these distributions in comparison to the Maxwellian case, we introduce the dimensionless parameter
\begin{equation}
\begin{aligned}
\frac{\delta R_i}{R_0} \equiv \frac{\langle\sigma v\rangle_i-\langle\sigma v\rangle_{\mathrm{MB}}}{\langle\sigma v\rangle_{\mathrm{MB}}}  = \frac{\int_0^{+\infty}\left[p_i(E)-p_{\mathrm{MB}}(E)\right] T(E) E^{-1 / 2} d E}{\int_0^{+\infty}p_{\mathrm{MB}}(E) T(E) E^{-1 / 2} d E},
\end{aligned}
\end{equation}
representing the relative deviation from the Maxwellian reactivity. As initially observed in Ref. \cite{Onofrio1}, if $T(E)$ were to scale exactly as $E^{1/2}$ across the entire energy range, the resulting difference would be zero, as both distributions are normalized to unity. However, because the tunneling coefficient is bounded by unity, such scaling cannot practically lead to breakeven. 
A more realistic expression for the energy-dependent tunneling coefficient is given by

\begin{equation}
T(E)= \begin{cases}\left(E / E_0\right)^\alpha, & E<E_0 \\ 1, & E>E_0 .\end{cases}
\end{equation}
As far as we know, this prototype model was initially proposed in Ref. \cite{Onofrio1}. In this context, the parameter $\alpha$ ($0 \leq \alpha < \infty$) serves as a ``convexity'' parameter. Specifically, for energies below $E_0$, the tunneling coefficient $T(E)$ is convex when $\alpha > 1$ and concave when $\alpha < 1$. When $T(E)$ is convex, the contribution to the integrated tunneling probability (\ref{R}) arises from both the high-energy and low-energy regions of the distribution. In contrast, when $T(E)$ is concave, the predominant contribution comes from the intermediate energy range.

For the Maxwellian case, the reactivity (\ref{R}) can be expressed explicitly as
\begin{equation}
R_0= \frac{  \hbar^2}{{2} m^{3 / 2}} \sqrt{\frac{{\pi}}{\langle E \rangle}} \left [{\Gamma\left(0, \frac{E_0}{2 \langle E \rangle}\right) +  \left(\frac{2\langle E \rangle}{E_0}\right)^{\alpha} \left(\Gamma(\alpha) - \Gamma\left(\alpha, \frac{E_0}{2 \langle E \rangle}\right)\right)}\right ],
\end{equation}
where $\Gamma(a)$ is the Euler Gamma function and
\begin{equation}
\Gamma(a, z): = \int_z^\infty t^{a-1} e^{-t} \, dt
\end{equation}
is the upper incomplete Gamma function.

\begin{figure*}
\centering
\begin{minipage}[t]{0.315\linewidth}
\includegraphics[width=1\linewidth]{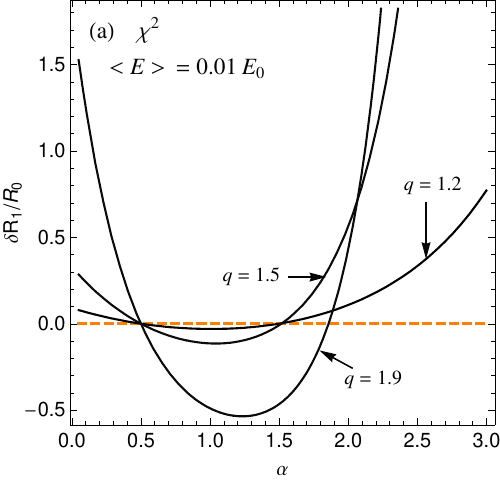}
\end{minipage}
\begin{minipage}[t]{0.3\linewidth}
\includegraphics[width=1\linewidth]{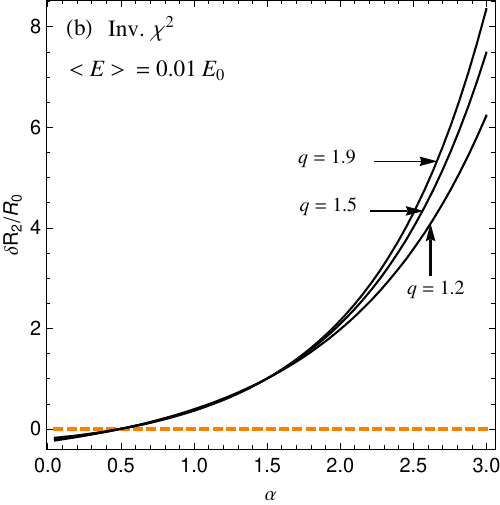}
\end{minipage}
\begin{minipage}[t]{0.3\linewidth}
\includegraphics[width=1\linewidth]{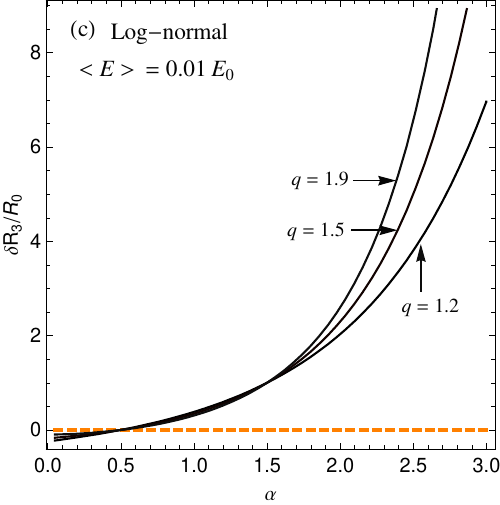}
\end{minipage}
\begin{minipage}[t]{0.3\linewidth}
\includegraphics[width=1\linewidth]{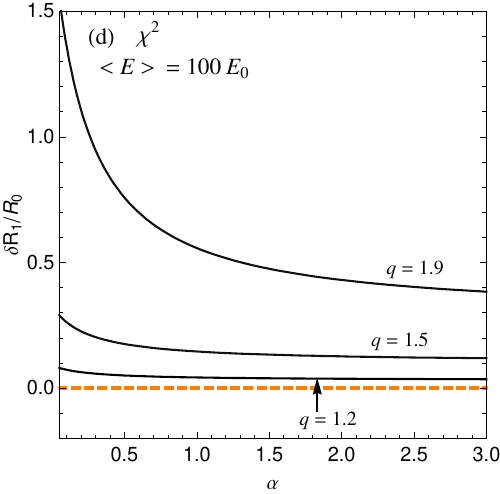}
\end{minipage}
\begin{minipage}[t]{0.3\linewidth}
\includegraphics[width=1\linewidth]{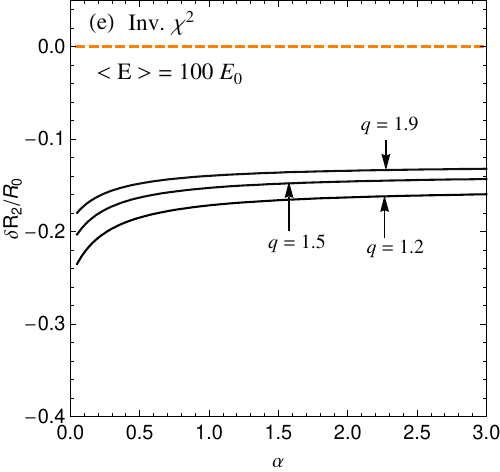}
\end{minipage}
\begin{minipage}[t]{0.3\linewidth}
\includegraphics[width=1\linewidth]{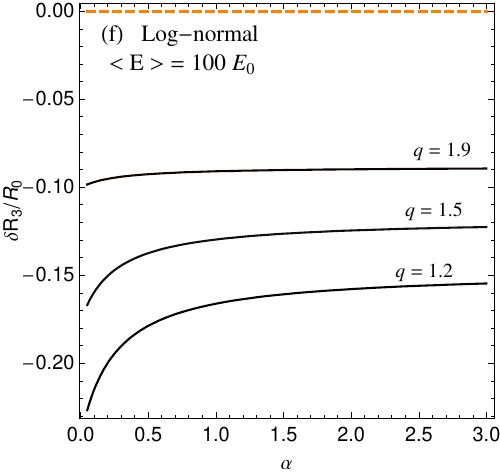}
\end{minipage}
\caption{Relative deviation from the Maxwellian reactivity, $\delta R_i / R_{{0}}$, as a function of the exponent of the tunneling coefficient, $\alpha$, for the three universality classes of superstatistics with various values of $q := \langle \beta^2 \rangle / \beta_0^2$. The upper panel corresponds to $\langle E \rangle = 0.01 E_0$, while the lower panel corresponds to $\langle E \rangle = 100 E_0$.} \label{Fig4}
\end{figure*}

For the three universality classes of superstatistics, we are unable to obtain a closed-form expression for the reactivities, so we compute them numerically. The results are shown in Fig. \ref{Fig4}, where we examine two distinct regimes based on the ratio between $\langle E \rangle$ and $E_0$: (i) the low average energy regime ($\langle E \rangle = 0.01 E_0$) and (ii) the high average energy regime ($\langle E \rangle = 100 E_0$). In both regimes, the behavior of the $\chi^2$ class (corresponding to Tsallis thermostatistics) is particularly distinctive. 

In the low average energy regime (upper panel), there is a range of $\alpha$ values where the reactivity exceeds that of the Maxwellian case. For the $\chi^2$ class, this occurs at both low and high values of $\alpha$, with the specific values depending on $q$ and the ratio $\langle E \rangle / E_0$. In contrast, for the other two classes of superstatistics, a threshold value of $\alpha$ exists, above which the reactivity significantly surpasses that of the Maxwellian distribution. In the high average energy regime (lower panel), the $\chi^2$ class shows a small enhancement for all values of $\alpha$, while the reactivity for the other two classes remains below that of the Maxwellian case, although it increases as $q$ deviates from unity. This particular behavior of the $\chi^2$ class can be attributed to its enhanced low-energy contribution compared to the Maxwellian case (see inset in Fig. \ref{Fig3}(a)).

As discussed in Sec. \ref{SecII}, regardless of the fluctuation class, superstatistical distributions exhibit universal behavior under small fluctuations, characterized by quadratic corrections to the Maxwellian case [see Eq. (\ref{al})]. It is therefore instructive to examine how this scenario influences the reactivity. In this case, the reactivity can be expressed explicitly as

\begin{equation}
 R= \frac{ \sqrt{\pi } \hbar^2}{ m^{3 / 2}}   \frac{\left [\zeta_1 + 
 \left(\frac{2 \langle E \rangle}{E_0}\right)^{\alpha} 
(\zeta_2 +  2 (q-1) \zeta_3  )
\right]}{2 + 3 (q-1) }
\end{equation}
where we have defined the following auxiliary functions

\begin{equation}
\begin{aligned}
 \zeta_1 (\langle E \rangle/ E_0, q):= &  { (q-1)(2  + E_0/ \langle E \rangle)e^{-\frac{E_0}{2 \langle E \rangle}} + 
        \Gamma\left(0, \frac{E_0}{2 \langle E \rangle}\right)}, \\
      \zeta_2 (\langle E \rangle/ E_0, \alpha):=&  \Gamma(\alpha) - \Gamma\left(\alpha, \frac{E_0}{2 \langle E \rangle}\right),\\
      \zeta_3 (\langle E \rangle/ E_0, \alpha):=& \Gamma(2 + \alpha) - \Gamma\left(2 + \alpha, \frac{E_0}{2 \langle E \rangle}\right).
\end{aligned}
\end{equation}

\begin{figure*}
\centering
\includegraphics[width=0.4\linewidth]{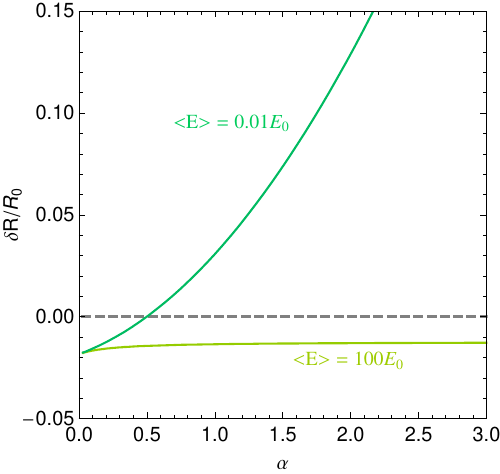}
\caption{Relative deviation of the reactivity, $\delta R / R_0$, as a function of the tunneling exponent $\alpha$, for the universal distribution (\ref{al}) corresponding to small fluctuations, in both regimes, with $q = 1.05$.} \label{Fig5}
\end{figure*}

The corresponding relative deviation from the Maxwellian reactivity, $\delta R / R_0$, is shown in Fig. \ref{Fig5} for both regimes. The behavior is qualitatively similar to that observed for inverse-$\chi^2$ and log-normal superstatistics. In the low average energy regime, the reactivity exceeds that of the Maxwellian for sufficiently large values of $\alpha$. On the other hand, in the high average energy regime, the reactivity consistently remains below that of the Maxwellian distribution.

\section{Ionization and recombination rates in a plasma}\label{SecIV}

In this section, we examine the influence of nonequilibrium distributions in a more concrete three-dimensional situation, specifically focusing on ionization and recombination rates in a plasma. Since plasma environments often exhibit quasiequilibrium behavior and are effectively modeled through superstatistics \cite{Plasma1,Plasma2,splasma3}, it is important to investigate how the shape of the distribution function affects the rates of ionization, excitation, and direct radiative recombination. For ionization and recombination, the rates are given, respectively, by

\begin{equation}
\begin{aligned}
& S_c \equiv \left\langle\sigma_I(v) v\right\rangle_i=4 \pi \int_0^{\infty} \sigma_I(v) v^3 P_i(v) d v, \\
& \alpha_R \equiv \left\langle\sigma_R(v) v\right\rangle_i=4 \pi \int_0^{\infty} \sigma_R(v) v^3 P_i(v) d v .
\end{aligned}
\end{equation}
where $i=1,2,3$ correspond to $\chi^2$, inverse-$\chi^2$, and log-normal superstatistics, while $\sigma_I$ and $\sigma_r$ denote the cross sections for ionization and recombination, respectively.
For $\sigma_I$, an accurate estimate is given by the semi-empirical Lotz formula \cite{Lotz}

\begin{equation}\label{sI}
\sigma_I=(0.7) 4 \pi a_0^2\left[\frac{\chi_H}{\chi}\right]^2 \xi \frac{\ln (E / \chi)}{E / \chi} \quad (E \geq \chi),
\end{equation}
where $E$ represents the kinetic energy of the incident electron, $a_0$ is the Bohr radius, $\chi$ is the ionization potential of the ion, and $\chi_H$ is the ionization potential of hydrogen ($13.6$ eV). The parameter $\xi$ denotes the number of electrons in the subshell from which ionization occurs, assumed to be $1$ in this case.

On the other hand, the main characteristics of the direct radiative recombination cross section $\sigma_R$ for level $n$ are well represented by the hydrogenic cross section \cite{H}

\begin{equation}\label{sR}
\sigma_R=\frac{32 \pi}{3 \sqrt{3}} \alpha^3 a_0^2 \frac{n}{(E / \chi)(1+E / \chi)},
\end{equation}
where $\alpha$ is the fine-structure constant. For simplicity, we assume in our calculations recombination to the ground state, i.e., $n=1$.

Although refinements to these cross sections are possible (see, for example, Ref. \cite{Lcor}), the expressions in Eqs. (\ref{sI}) and (\ref{sR}) are commonly employed in the recent literature \cite{L0,L1,L2} and are generally accurate. The primary source of uncertainty lies in the numerical coefficients, whereas the energy dependence of the cross sections remains reliable, particularly in the high-energy limit.

\begin{figure*}
\centering
\begin{minipage}[t]{0.3\linewidth}
\includegraphics[width=1\linewidth]{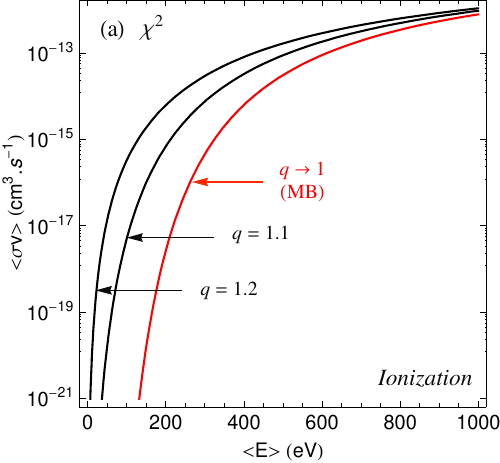}
\end{minipage}
\begin{minipage}[t]{0.3\linewidth}
\includegraphics[width=1\linewidth]{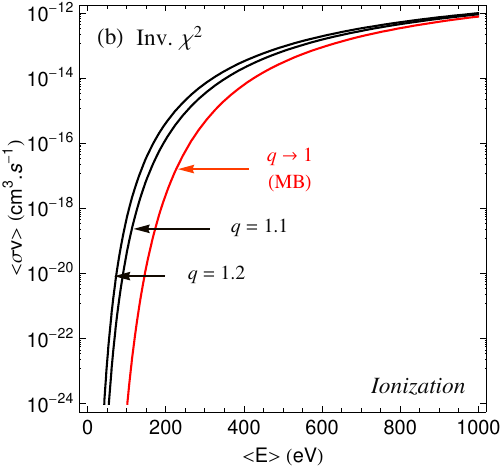}
\end{minipage}
\begin{minipage}[t]{0.3\linewidth}
\includegraphics[width=1\linewidth]{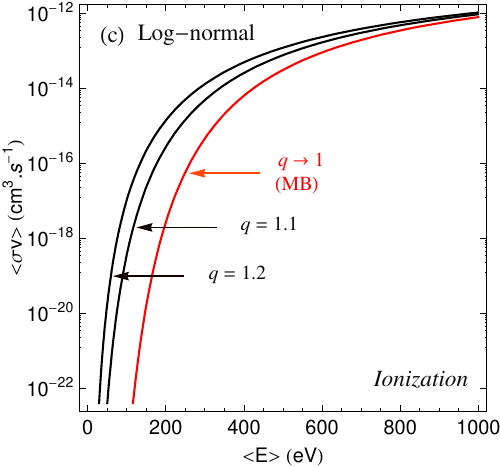}
\end{minipage}
\begin{minipage}[t]{0.3\linewidth}
\includegraphics[width=1\linewidth]{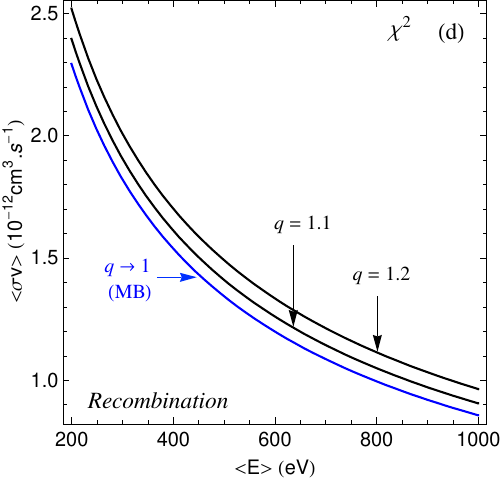}
\end{minipage}
\begin{minipage}[t]{0.3\linewidth}
\includegraphics[width=1\linewidth]{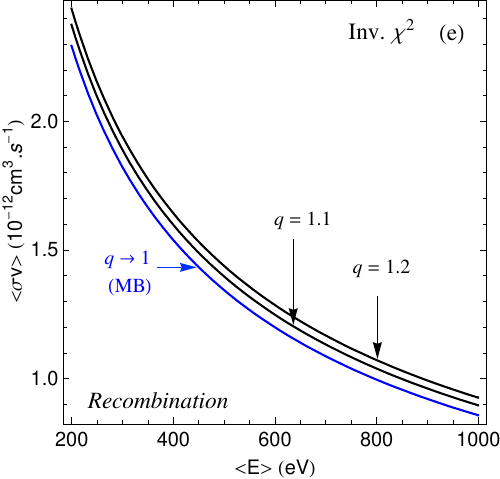}
\end{minipage}
\begin{minipage}[t]{0.3\linewidth}
\includegraphics[width=1\linewidth]{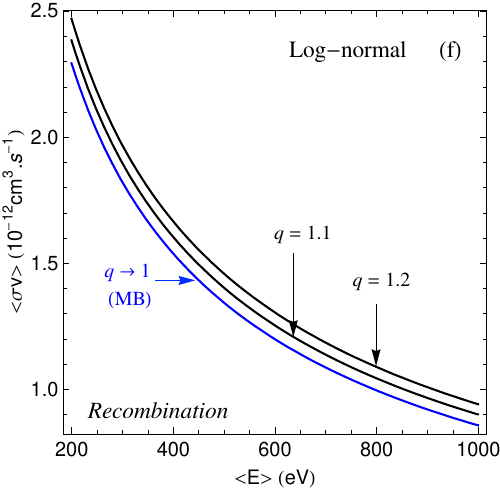}
\end{minipage}
\caption{Rates of collisional ionization (upper panel) and direct radiative recombination (lower panel) as a function of the average energy $\langle E \rangle$ (eV) for the three universality classes of superstatistics, with varying values of $q := \langle \beta^2 \rangle / \beta_0^2$. The ionization potential is set at $\chi = 2000 \text{eV}$.} \label{Fig6}
\end{figure*}

Using the three universality classes of superstatistics, we numerically calculated the associated ionization and recombination rates. Figure \ref{Fig6} displays these rates as a function of the average energy $\langle E \rangle$. For illustrative purposes, we set $\chi = 2000$ eV, applicable for example for He-like and H-like aluminum or M-shell gold ions \cite{PRA}. It is evident that the three classes of superstatistics contribute to an increase in both the ionization and recombination rates. For reasonable values of $q$, the increase in recombination rates remains modest, staying below 
$13\%$. However, the ionization rates may exhibit increases by several orders of magnitude relative to their Maxwellian equivalents, 
when the ionization potential $\chi$ is sufficiently high. This is due to the extended high-energy tails of superstatistical distributions, leading to a significant surplus of electrons with energies $\geq \chi$.

For lower ionization potentials, the effects are more subtle. This is illustrated in Table \ref{Table}, which presents the numerically computed ionization and recombination rates under typical solar wind conditions, where hydrogen dominates and the temperature is around $T = 10^5$ K. Solar wind velocity distributions are known to deviate from the Maxwellian distribution \cite{azerty3}, fitting well with both the $\chi^2$ and log-normal classes \cite{splasma3}. For the sake of illustration, we assume for \ref{Table} a moderate value of $q=1.2$. We observe that the increase in the ionization and recombination rates is marginal under these conditions.

\begin{table}[h!]
\centering
\begin{tabular}{ccccc}
\hline\hline
& Maxwellian& $\chi^2$ & Inverse $\chi^2$ & Log-normal \\ \hline
$\langle \sigma_I(v) v \rangle$ ($10^{-9} \text{cm}^3 / \text{s}$) & 2.78 & 3.16 & 3.11 & 3.15 \\ \hline
$\langle \sigma_R(v) v \rangle$ ($10^{-14} \text{cm}^3 / \text{s}$) & 5.99 & 6.78 & 6.50 & 6.61 \\ \hline\hline
\end{tabular}
\caption{Numerical values of ionization and recombination rates for $T=10^5$ K—characteristic of solar wind conditions—with $q=1.2$.}\label{Table}
\end{table}

\section{Conclusion}\label{SecV}


In this paper, we explored reaction rates in quasiequilibrium systems, which exhibit only local equilibrium. These systems are characterized by a hierarchical structure in their dynamics, allowing their statistical properties to be represented as a superposition of different statistics, i.e., superstatistics. This approach has proven effective in modeling diverse scenarios where reaction rates play a significant role, such as in plasmas \cite{Plasma2,splasma3}. We focused on the three universality classes of superstatistics, which correspond to three typical phenomenological mechanisms likely responsible for the emergence of temperature as a fluctuating variable (see Ref. \cite{CLT} for details). Supported by compelling empirical evidence, these classes constitute a reliable basis for understanding the statistical properties of quasiequilibrium systems. While our analysis centered on these specific universality classes, the framework is versatile enough to accommodate other scenarios, provided that a clear scale separation exists in the system's dynamics.

We examined an idealized one-dimensional model simulating tunneling processes like fusion, as proposed in Ref. \cite{Onofrio1}, and showed that the three classes of superstatistics can increase fusion reactivities under certain conditions. This finding extends recent results \cite{Onofrio1} on kappa distributions, characterized by power-law behavior in the high-energy limit, and bi-modal Maxwellian distributions, both of which can be interpreted as forms of superstatistics. Additionally, we explored a more concrete scenario to enable quantitative analysis: ionization and recombination rates in a plasma, using semi-empirical cross sections. In this case as well, we observed that these distributions, characteristic of quasiequilibrium systems, contribute to an increase in reaction rates.

This paper suggests potential directions for future research. In particular, extending the study to include quantum versions of the superstatistical distributions (as discussed in Ref. \cite{OurabahSR}), or their relativistic counterparts, could be a valuable next step. In fact, reactions occurring in plasmas sufficiently hot for relativistic effects to be non-negligible play an important role, for example, in studies of supernovae, X-ray sources, and other high-energy astrophysical phenomena \cite{Weaver}. 
Moreover, while superstatistical distributions appear to outperform Maxwellian distributions in enhancing fusion reactivities under certain conditions, their practical implementation in thermonuclear fusion remains an open question. Insights in this direction can be obtained experimentally through the spectroscopy of fusion reaction products, which may provide accurate assessments of the deviations from the Maxwellian distribution \cite{Crilly}.

\end{document}